# Analytical computation of bifurcation of orbits near collinear libration point in the restricted three-body problem

Mingpei Lin[1,*], Tong Luo[2,3,†] and Hayato Chiba[1,‡]

[1] Advanced Institute for Materials Research, Tohoku University, Sendai, 980-8577, Japan

[2] Beijing Institute of Control Engineering, Beijing, 100186, China

National key laboratory of space intelligent control, Beijing, 100186, China

**Abstract**: A unified analytical solution is presented for constructing the phase space near collinear libration points in the Circular Restricted Three-body Problem (CRTBP), encompassing Lissajous orbits and quasihalo orbits, their invariant manifolds, as well as transit and non-transit orbits. Traditional methods could only derive separate analytical solutions for the invariant manifolds of Lissajous orbits and halo orbits, falling short for the invariant manifolds of quasihalo orbits. By introducing a coupling coefficient $\eta$ and a bifurcation equation, a unified series solution for these orbits is systematically developed using a coupling-induced bifurcation mechanism and Lindstedt-Poincaré method. Analyzing the third-order bifurcation equation reveals bifurcation conditions for halo orbits, quasihalo orbits, and their invariant manifolds. Furthermore, new families of periodic orbits similar to halo orbits are discovered, and non-periodic/quasi-periodic orbits, such as transit orbits and non-transit orbits, are found to undergo bifurcations. When $\eta = 0$, the series solution describes Lissajous orbits and their invariant manifolds, transit, and non-transit orbits. As $\eta$ varies from zero to non-zero values, the solution seamlessly transitions to describe quasihalo orbits and their invariant manifolds, as well as newly bifurcated transit and non-transit orbits. This unified analytical framework provides a more comprehensive understanding of the complex phase space structures near collinear libration points in the CRTBP.

**Keywords**: Restricted three-body problem; Coupling coefficient; Bifurcation equation; Phase space; Invariant manifolds; High-order series solution

## 1. Introduction

The Circular Restricted Three-body Problem (RTBP) stands as a classic dynamical model for exploring the motion of asteroids or satellites under the gravitational attraction of two primaries [1−4]. The CRTBP has five equilibrium points called the Lagrange points $L_i$ ($i = 1, 2, …, 5$), including three so-called collinear libration points and two triangular libration points. The linear behavior of three collinear libration points is of the type center × center × saddle. The phase space near these collinear libration points is intricately structured, featuring a diverse array of orbits, including one- and two-dimensional invariant tori, hyperbolic manifolds corresponding to libration points and periodic/quasi-periodic orbits, as well as transit and non-transit trajectories [5−7]. These libration orbits play a crucial role in many space science missions [8−10].

[*] lin.mingpei.d2@tohoku.ac.jp
[†] luotong@buaa.edu.cn
[‡] Corresponding author, hchiba@tohoku.ac.jp

Considerable attention has been devoted to the investigation of transfer orbits within the CRTBP. In its early stages, numerical trial-and-error methodologies were predominant for designing transfer orbits, demanding a profound understanding of space mission dynamics, and proving to be a time-intensive endeavor [11,12]. Gómez et al. [13,14] pioneered the application of invariant manifold theory to aid in designing transfer orbits in the CRTBP. Howell et al. [3] employed these methods, coupled with differential correction techniques to develop a powerful tool for designing transfer orbits from Earth Parking orbit to halo orbits in the Sun-Earth system. Obtaining analytical initial solutions for these orbits is crucial for identifying suitable target mission orbits and transfer trajectories. Leveraging ad hoc algebraic manipulations, Jorba et al. [15] achieved semi-analytical solutions of Lissajous orbits and halo orbits up to order 35. Building upon this, Masdemont [16] presented semi-analytical solutions for the invariant manifolds of Lissajous and halo orbits. However, a notable gap remained in their inability to provide a semi-analytical solution for invariant manifolds of quasihalo orbits.

Lin and Chiba [17] presented a coupling-induced bifurcation mechanism to explain how quasihalo orbits bifurcate from Lissajous orbits. They proposed that coupling between different degrees of freedom in dynamical systems is the underlying cause of this bifurcation. Through the introduction of coupling coefficients and bifurcation equations, they successfully elucidated the generation of halo and quasihalo orbits, providing a unified high-order analytical solution for the center manifolds of collinear libration points. In this paper, we extend the concept that coupling induces bifurcations to the entire phase space in the CRTBP. By utilizing the Lindstedt-Poincaré method and the concept of coupling coefficients, a complete three-parameter bifurcation equation is derived, with each parameter corresponding to one degree of freedom in the CRTBP. Analyzing this bifurcation equation reveals that bifurcations occur not only for invariant tori but also for invariant manifolds and general transit and non-transit orbits in phase space. This results in a unified high-order analytical solution for describing the phase space near the libration points. As a special case, the high-order analytical solution for the invariant manifold of quasihalo orbits can be naturally obtained as a result.

The subsequent sections of this paper are organized as follows. The dynamical model of the CRTBP is introduced in Section 2. A high-order analytical solution for phase space near collinear libration points in the CRTBP is constructed in Section 3. Bifurcation analysis and numerical results are presented in Section 4. Finally, Section 5 offers concluding remarks.

## 2. Dynamical model

The CRTBP serves as a good approximation for studying the motion of infinitesimal particles, such as asteroids or spacecraft, under the gravitational influence of two primary celestial bodies. In this model, two primaries rotating around their common center of mass in a circular orbit. The motion of the particle is described in a synodic coordinate system, where the origin is established at the centroid of the two primaries. The $X$-axis spans from the smaller primary to larger primary, the $Z$-axis is perpendicular to the plane of the circular orbit, directed positively along the angular momentum vector, and the $Y$-axis completes a right-hand triad. Let $\mu$ denote the ratio of the mass of the smaller celestial body to the sum of the masses of the two bodies, given by $\mu = m_2 / (m_1 + m_2)$. The normalized coordinates for the smaller and larger celestial bodies are $(-\mu, 0, 0)$ and $(1-\mu, 0, 0)$, respectively. The governing differential equation dictating the particle's motion in the normalized synodic coordinate system is given by [15]

$$\ddot{X} - 2\dot{Y} = \frac{\partial \Omega}{\partial X}$$
$$\ddot{Y} + 2\dot{X} = \frac{\partial \Omega}{\partial Y} \qquad (1)$$
$$\ddot{Z} = \frac{\partial \Omega}{\partial Z}$$

with

$$\Omega(X,Y,Z) = \frac{1}{2}\left(X^2 + Y^2\right) + \frac{1-\mu}{r_1} + \frac{\mu}{r_2} + \frac{1}{2}\mu(1-\mu). \qquad (2)$$

Here $r_1$ and $r_2$ denote the distances from the particle to the smaller primary and larger primary, respectively,

$$r_1^2 = (X+\mu)^2 + Y^2 + Z^2$$
$$r_2^2 = (X-1+\mu)^2 + Y^2 + Z^2. \qquad (3)$$

Equation (1) has a Jacobi integral $C = 2\Omega - (\dot{X}^2 + \dot{Y}^2 + \dot{Z}^2)$ and five equilibrium points, with our focus directed towards the three collinear libration points, namely $L_1$, $L_2$ and $L_3$. Let $\gamma_i$ ($i = 1, 2, 3$) represent the distance from $L_i$ to the closet primary. They are determined by the unique positive root of the Euler quantic equation [15],

$$\gamma_i^5 \mp (3-\mu)\gamma_i^4 + (3-2\mu)\gamma_i^3 - \mu\gamma_i^2 \pm 2\mu\gamma_i - \mu = 0, \quad i = 1, 2$$
$$\gamma_i^5 + (2+\mu)\gamma_i^4 + (1+2\mu)\gamma_i^3 - (1-\mu)\gamma_i^2 - 2(1-\mu)\gamma_i - (1-\mu) = 0, \quad i = 3. \qquad (4)$$

When focusing on phase space near the chosen libration point $L_i$, a coordinate transformation is implemented to relocate the origin of synodic coordinate system to the libration point $L_i$,

$$X = -\gamma_i x + \mu - 1 + \gamma_i, Y = -\gamma_i y, Z = \gamma_i z, \quad i = 1, 2$$
$$X = \gamma_i x + \mu + \gamma_i, Y = \gamma_i y, Z = \gamma_i z, \quad i = 3. \qquad (5)$$

Then, (1) is rewritten as

$$\ddot{x} - 2\dot{y} = \frac{\partial \Omega}{\partial x}$$
$$\ddot{y} + 2\dot{x} = \frac{\partial \Omega}{\partial y} \quad (6)$$
$$\ddot{z} = \frac{\partial \Omega}{\partial z}$$

with

$$\Omega(x,y,z) = \frac{1}{2}\left((\mu - 1 \mp \gamma(x-1))^2 + \gamma^2 y^2\right) + \frac{1-\mu}{r_1} + \frac{\mu}{r_1} + \frac{1}{2}\mu(1-\mu). \quad (7)$$

To facilitate the construction of a unified analytical solution for the phase space near collinear libration point in the subsequent section, the right-hand side of (6) is expanded in a power series using Legendre polynomials [15, 16],

$$\ddot{x} - 2\dot{y} - (1+2c_2)x = \frac{\partial}{\partial x}\sum_{n\geq 3} c_n \rho^n P_n\left(\frac{x}{\rho}\right)$$
$$\ddot{y} + 2\dot{x} + (c_2 - 1)y = \frac{\partial}{\partial y}\sum_{n\geq 3} c_n \rho^n P_n\left(\frac{y}{\rho}\right) \quad (8)$$
$$\ddot{z} + c_2 z = \frac{\partial}{\partial z}\sum_{n\geq 3} c_n \rho^n P_n\left(\frac{z}{\rho}\right)$$

where $\rho = x^2 + y^2 + z^2$, $P_n$ represents Legendre polynomials, and the constant coefficients $c_n(\mu)$ is determined by

$$c_n(\mu) = \frac{1}{\gamma_i^3}\left((\pm 1)^n + (-1)^n \frac{(1-\mu)\gamma_i^{n+1}}{(1\mp \gamma)^{n+1}}\right), \quad \text{for } L_i, i=1,2$$
$$c_n(\mu) = \frac{(-1)^n}{\gamma_i^3}\left(1-\mu + \frac{\mu\gamma_i^{n+1}}{(1+\gamma)^{n+1}}\right), \quad \text{for } L_i, i=3. \quad (9)$$

## 3. Analytical construction of phase space near collinear libration points

In this section, we introduce the concept of coupling coefficients and employ the Lindstedt–Poincaré method to construct a semi-analytical solution for phase space near collinear libration points in the RTBP. The process initiates with a linear solution that considers the coupling effect between in-plane and out-of-plane motion. The Lindstedt–Poincaré method is then applied iteratively to refine the relationship between frequency and amplitude. Through stepwise iteration of the known low-order solution, a higher-order series solution and a polynomial bifurcation equation are obtained.

### 3.1 Form of the analytical solution

First, the linear solution for the phase space near collinear libration points in the RTBP is derived from the linear component of (8),

$$\ddot{x} - 2\dot{y} - (1+2c_2)x = 0$$
$$\ddot{y} + 2\dot{x} + (c_2 - 1)y = 0 \quad (10)$$
$$\ddot{z} + c_2 z = 0.$$

It is known that quasi-periodic solution of (10) incorporates both in-plane and out-of-plane motions with different frequencies, while the hyperbolic exponential solution comprises stable and unstable components with opposites exponents. Thus, the complete solution of (10) takes the form:

$$x(t) = \alpha_1 \cos(\omega_0 t + \varphi_1) + \alpha_3 e^{\lambda_0 t} + \alpha_4 e^{-\lambda_0 t}$$
$$y(t) = \kappa_1 \alpha_1 \sin(\omega_0 t + \varphi_1) + \kappa_2 \alpha_3 e^{\lambda_0 t} - \kappa_2 \alpha_4 e^{-\lambda_0 t} \qquad (11)$$
$$z(t) = \alpha_2 \cos(\nu_0 t + \varphi_2)$$

where

$$\omega_0 = \sqrt{\frac{2 - c_2 + \sqrt{9c_2^2 - 8c_2}}{2}}, \lambda_0 = \sqrt{\frac{c_2 - 2 + \sqrt{9c_2^2 - 8c_2}}{2}}, \nu_0 = \sqrt{c_2}$$
$$\kappa_1 = -\frac{\omega_0^2 + 1 + c_2}{2\omega_0}, \kappa_2 = \frac{\lambda_0^2 - 1 - c_2}{2\lambda_0}. \qquad (12)$$

$\alpha_1$ and $\alpha_2$ are the in-plane and out-of-plane amplitudes in center part, respectively. $\varphi_1$ and $\varphi_2$ are the corresponding phases, respectively. $\alpha_3$ and $\alpha_4$ are amplitudes of stable and unstable components in hyperbolic part.

As mentioned in Introduction section, halo/quasihalo orbits and their invariant manifolds bifurcate from the planar Lyapunov/quasihalo orbits and their invariant manifolds due to the nonlinear coupling of in-plane and out-of-plane motions in CRTBP. Considering this nonlinear coupling, we modify the equations of motion (8) by introducing the product of a coupling coefficient $\eta$, a correction factor $\Delta$ and $x$ to the third equation,

$$\ddot{x} - 2\dot{y} - (1 + 2c_2)x = \frac{\partial}{\partial x} \sum_{n \geq 3} c_n \rho^n P_n\left(\frac{x}{\rho}\right)$$
$$\ddot{y} + 2\dot{x} + (c_2 - 1)y = \frac{\partial}{\partial y} \sum_{n \geq 3} c_n \rho^n P_n\left(\frac{y}{\rho}\right) \qquad (13)$$
$$\ddot{z} + c_2 z = \frac{\partial}{\partial z} \sum_{n \geq 3} c_n \rho^n P_n\left(\frac{z}{\rho}\right) + \eta \Delta x$$

where $\eta$ quantifies the extent of coupling between in-plane and out-of-plane motions, and $\Delta$ represents the correction factor satisfying a so-called bifurcation equation $\Delta = 0$. As a result, the modified linear part of (10) becomes:

$$\ddot{x} - 2\dot{y} - (1 + 2c_2)x = 0$$
$$\ddot{y} + 2\dot{x} + (c_2 - 1)y = 0 \qquad (14)$$
$$\ddot{z} - \eta d_{0000} x + c_2 z = 0.$$

The linear part of solution for (14) can be obtained as

$$x(t) = \alpha_1 \cos(\omega_0 t + \varphi_1) + \alpha_3 e^{\lambda_0 t} + \alpha_4 e^{-\lambda_0 t}$$
$$y(t) = \kappa_1 \alpha_1 \sin(\omega_0 t + \varphi_1) + \kappa_2 \alpha_3 e^{\lambda_0 t} - \kappa_2 \alpha_4 e^{-\lambda_0 t} \qquad (15)$$
$$z(t) = \alpha_2 \cos(\nu_0 t + \varphi_2) + \eta \alpha_1 \cos(\omega_0 t + \varphi_1) + \eta \kappa_3 \alpha_3 e^{\lambda_0 t} + \eta \kappa_3 \alpha_4 e^{-\lambda_0 t}.$$

with $d_{0000} = v_0^2 - \omega_0^2$, $\kappa_3 = \dfrac{v_0^2 - \omega_0^2}{v_0^2 + \lambda_0^2}$. It can be observed that for the first-order solution, $\Delta = d_{0000} \neq 0$, indicating the absence of coupling between in-plane and out-of-plane motions. Quasihalo orbits and their invariant manifolds emerge only when the coupling coefficient $\eta$ is not zero.

Upon considering the nonlinear terms in equation (14), the high-order solution is expressed as formal expansions in powers of $\alpha_i$ ($i = 1,2,3,4$):

$$x(t) = \sum \left[ x_{ijkm}^{pq} \cos(p\theta_1 + q\theta_2) + \overline{x}_{ijkm}^{pq} \sin(p\theta_1 + q\theta_2) \right] e^{(k-m)\theta_3} \alpha_1^i \alpha_2^j \alpha_3^k \alpha_4^m$$
$$y(t) = \sum \left[ y_{ijkm}^{pq} \cos(p\theta_1 + q\theta_2) + \overline{y}_{ijkm}^{pq} \sin(p\theta_1 + q\theta_2) \right] e^{(k-m)\theta_3} \alpha_1^i \alpha_2^j \alpha_3^k \alpha_4^m \quad (16)$$
$$z(t) = \sum \left[ z_{ijkm}^{pq} \cos(p\theta_1 + q\theta_2) + \overline{z}_{ijkm}^{pq} \sin(p\theta_1 + q\theta_2) \right] e^{(k-m)\theta_3} \alpha_1^i \alpha_2^j \alpha_3^k \alpha_4^m$$

where $\theta_1 = \omega t + \varphi_1$, $\theta_2 = vt + \varphi_2$, $\theta_3 = \lambda t$. Due to the nonlinear terms, the frequencies should also be expanded in the power series of $\alpha_i$ ($i = 1,2,3,4$),

$$\omega = \sum \omega_{ijkm} \alpha_1^i \alpha_2^j \alpha_3^k \alpha_4^m, \quad v = \sum v_{ijkm} \alpha_1^i \alpha_2^j \alpha_3^k \alpha_4^m, \quad \lambda = \sum \lambda_{ijkm} \alpha_1^i \alpha_2^j \alpha_3^k \alpha_4^m, \quad (17)$$

$\Delta$ is expanded as a series with the amplitudes $\alpha_i$ ($i = 1,2,3,4$), i.e.,

$$\Delta = \sum d_{ijkm} \alpha_1^i \alpha_2^j \alpha_3^k \alpha_4^m. \quad (18)$$

Here ($x_{ijkm}^{pq}, \overline{x}_{ijkm}^{pq}$), ($y_{ijkm}^{pq}, \overline{y}_{ijkm}^{pq}$) and ($z_{ijkm}^{pq}, \overline{z}_{ijkm}^{pq}$) is the paired coefficients corresponding to the coordinate series $x$, $y$ and $z$, respectively. $\omega_{ijkm}$, $v_{ijkm}$, $\lambda_{ijkm}$ and $d_{ijkm}$ are coefficients corresponding to the frequency series $\omega$, $v$, $\lambda$ and $\Delta$, respectively.

*Remark 1.* To ensure (16) is the solution for the original equation, we have the constraint $\eta \Delta = 0$, which provide the implicit relationship between $\eta$ and $\alpha_i$, i.e., $\eta = \eta(\alpha_1, \alpha_2, \alpha_3, \alpha_4)$. When $\alpha_1 \neq 0$, $\alpha_2 \neq 0$, $\eta = 0$ and $\Delta \neq 0$, the solution (16) represents Lissajous orbits ($\alpha_3 = \alpha_4 = 0$), their stable ($\alpha_3 = 0$, $\alpha_4 \neq 0$) and unstable manifolds ($\alpha_3 \neq 0, \alpha_4 = 0$), transit ($\alpha_3 \times \alpha_4 < 0$) and non-transit orbits ($\alpha_3 \times \alpha_4 > 0$). Especially, if $\alpha_2 = 0$, it is planar Lyapunov orbits, if $\alpha_1 = 0$, it is vertical Lyapunov orbits; When $\alpha_1 \neq 0$, $\alpha_2 \neq 0$, $\eta = 0$ and $\Delta = 0$, Lissajous orbits and their corresponding invariant orbits undergo a bifurcation precisely; When $\alpha_1 \neq 0$, $\alpha_2 \neq 0$, $\eta \neq 0$ and $\Delta = 0$, solution (16) represent quasihalo orbit ($\alpha_3 = \alpha_4 = 0$), their stable ($\alpha_3 = 0, \alpha_4 \neq 0$) and unstable manifolds ($\alpha_3 \neq 0, \alpha_4 = 0$), transit ($\alpha_3 \times \alpha_4 < 0$) and non-transit orbits ($\alpha_3 \times \alpha_4 > 0$). Especially, if $\eta > 0$ ($\eta < 0$), $\alpha_2 = 0$, it is northern (southern) halo orbits. For the case $\alpha_1 = 0$, $\alpha_2 = 0$, there are similar results, if $\eta = 0$ and $\Delta \neq 0$, the solution (16) represents stable ($\alpha_3 = 0, \alpha_4 \neq 0$) and unstable manifolds ($\alpha_3 \neq 0, \alpha_4 = 0$) transit ($\alpha_3 \times \alpha_4 < 0$) and non-transit orbits ($\alpha_3 \times \alpha_4 > 0$) of libration points (LPs); if $\eta \neq 0$ and $\Delta = 0$, these invariant and transit/non-transit orbits bifurcate new types of orbits, which will be analyzed in the next section. To sum up, solution (16) completely describes the phase space near collinear libration points, shown in Table 1.

Table 1 Classification of the classical orbits near collinear Libration points

| ($\alpha_1, \alpha_2$) \ ($\alpha_3, \alpha_4$) | | Center manifolds (0, 0) | Unstable manifolds ($\neq 0$, 0) | Stable manifolds (0, $\neq 0$) | Non-transit orbits ($\alpha_3 \times \alpha_4 > 0$) | Transit orbits ($\alpha_3 \times \alpha_4 < 0$) |
|---|---|---|---|---|---|---|
| (0, 0) | $\eta=0$ | Libration points | Invariant manifolds of LPs | | Orbits near LPs | |
| | $\eta \neq 0$ | | | | Newly bifurcated orbits near LPs | |
| (0, $\neq 0$) | $\eta=0$ | Vertical Lyapunov orbits and their invariant manifolds | | | $\eta = 0$: corresponding non-transit and transit orbits. $\eta \neq 0$: type of bifurcation depends on the value of the amplitudes. | |
| ($\neq 0$, 0) | $\eta=0$ | Planar Lyapunov orbits and their invariant manifolds | | | | |
| | $\eta \neq 0$ | Halo orbits and their invariant manifolds | | | | |
| ($\neq 0, \neq 0$) | $\eta=0$ | Lissajous orbits and their invariant manifolds | | | | |
| | $\eta \neq 0$ | Quasihalo orbits and their invariant manifolds | | | | |

*Remark 2*. Owing to the inclusion of exponential terms for the hyperbolic component, the series of $x$, $y$, and $z$ terms no longer follow the form of cosine, sine, and cosine format [15]. In (16) and (17), $i, j, k$, and $j \in \mathbb{N}$, $p$ and $q \in \mathbb{Z}$. Due to the symmetry of the CRTBP, $p$ and $q$ have the same parity as ($i + j$) and ($k + m$), respectively. Furthermore, leveraging the symmetry properties of sine and cosine functions allows the assumption that $p \geq 0$, and $q \geq 0$ when $p = 0$. The series of $\omega$ and $\upsilon$ only include those even items. These facts are useful for saving computing storage and time.

**3.2 Solving for undetermined coefficients**

This subsection proceeds with the computation of the coefficients ($x_{ijkm}^{pq}, \bar{x}_{ijkm}^{pq}$), ($y_{ijkm}^{pq}, \bar{y}_{ijkm}^{pq}$), ($z_{ijkm}^{pq}, \bar{z}_{ijkm}^{pq}$) in (16), $\omega_{ijkm}$, $\upsilon_{ijkm}$, $\lambda_{ijkm}$ in (17), and $d_{ijkm}$ in (18) up to finite order $n$ using Lindstedt–Poincaré method. Starting with the linear solution (15), subsequent coefficients for higher orders are obtained through iterative calculations.

Upon reaching coefficients up to order $n - 1$, $x(t)$, $y(t)$, and $z(t)$ are determined up to order $n - 1$, $\omega, \upsilon, \lambda$, and $\Delta$ are determined up to order $n - 2$, we substitute these into the right side of (13), resulting in three power series $p$, $q$, and $r$ up to $n$ order. Our primary focus is on those $n$-order terms. The $n$-order terms of $p$, $q$, and $r$ are denoted by ($r_{ijkm}^{pq}, \bar{r}_{ijkm}^{pq}$), ($s_{ijkm}^{pq}, \bar{s}_{ijkm}^{pq}$) and ($t_{ijkm}^{pq}, \bar{t}_{ijkm}^{pq}$) ($i + j + k + m = n$) respectively, and the $n$-order terms of $\omega, \upsilon, \lambda$, and $\Delta$ are denoted by $\omega_{ijkm}$, $\upsilon_{ijkm}$, $\lambda_{ijkm}$ and $d_{ijkm}$ respectively.

Subsequently, the composition of $n$-order terms on the left side of (13) is analyzed. Based on (16), the derivatives of variable $x$ can be given by

$$\dot{x} = \omega \frac{\partial x}{\partial \theta_1} + \upsilon \frac{\partial x}{\partial \theta_2} + \lambda \frac{\partial x}{\partial \theta_3}$$

$$\ddot{x} = \omega^2 \frac{\partial^2 x}{\partial \theta_1^2} + \upsilon^2 \frac{\partial^2 x}{\partial \theta_2^2} + \lambda^2 \frac{\partial^2 x}{\partial \theta_3^2} + 2\omega\upsilon \frac{\partial^2 x}{\partial \theta_1 \partial \theta_2} + 2\omega\lambda \frac{\partial^2 x}{\partial \theta_1 \partial \theta_3} + 2\upsilon\lambda \frac{\partial^2 x}{\partial \theta_2 \partial \theta_3}. \quad (19)$$

The derivatives of $y$ can be similarly obtained. Let $fg$ represents the multiplication of frequency or coupling factor ($\omega, \upsilon, \lambda$, and $\Delta$) and the coordinate variables ($x, y,$ and $z$). $(ijkm)_f$ and $(ijkm)_g$ denote their corresponding order. Our goal is to identify and classify those $n$-order $fg$ satisfying $(ikmj)_f + (ijkm)_g = n$. When $(ij)_f$ is 0 or $n - 1$, the corresponding $(ij)_g$ is $n$ or 1, making $fg$ an unknown term to be solved. Conversely, when $(ij)_f = 1, 2, …, n-2$, $fg$ is a known term, to be moved to the right side of (13). Table 2 succinctly outlines the unknown first derivatives of $x$ and $y$ and the product of $\Delta$ and $x$, where $\delta_{ij}$ denotes Kronecker function. Analogously, Table 3 provides a summary for the unknown second derivatives of $x$, $y$, and $z$.

Table 2 Derivatives of $x$ and $y$ with respect to time and $\Delta x$

| | |
|---|---|
| $\omega \dfrac{\partial x}{\partial \theta_1} \to \begin{cases} \omega_0(p\bar{x}_{ijkm}^{pq}, -px_{ijkm}^{pq}) \\ (0, -\omega_{i-1jkm}\delta_{1p}\delta_{0q}\delta_{km}), \end{cases}$ | $\omega \dfrac{\partial y}{\partial \theta_1} \to \begin{cases} \omega_0(p\bar{y}_{ijkm}^{pq}, -py_{ijkm}^{pq}) \\ (\omega_{i-1jkm}\kappa_1\delta_{1p}\delta_{0q}\delta_{km}, 0) \end{cases}$ |
| $v \dfrac{\partial x}{\partial \theta_2} \to v_0(q\bar{x}_{ijkm}^{pq}, -qx_{ijkm}^{pq}),$ | $v \dfrac{\partial y}{\partial \theta_2} \to v_0(q\bar{y}_{ijkm}^{pq}, -qy_{ijkm}^{pq})$ |
| $\lambda \dfrac{\partial x}{\partial \theta_3} \to \begin{cases} \lambda_0((k-m)x_{ijkm}^{pq}, (k-m)\bar{x}_{ijkm}^{pq}) \\ (\lambda_{ijmm}\delta_{0p}\delta_{0q}\delta_{k-1m} - \lambda_{ijkk}\delta_{0p}\delta_{0q}\delta_{km-1}, 0), \end{cases}$ | $\lambda \dfrac{\partial y}{\partial \theta_3} \to \begin{cases} \lambda_0((k-m)y_{ijkm}^{pq}, (k-m)\bar{y}_{ijkm}^{pq}) \\ (\kappa_2\lambda_{ijmm}\delta_{0p}\delta_{0q}\delta_{k-1m} + \kappa_2\lambda_{ijkk}\delta_{0p}\delta_{0q}\delta_{km-1}) \end{cases}$ |
| $\Delta x \to \begin{cases} d_0(x_{ijkm}^{pq}, \bar{x}_{ijkm}^{pq}) \\ (d_{i-1jkm}\delta_{1p}\delta_{0q}\delta_{km}, 0) \\ (d_{ijmm}\delta_{0p}\delta_{0q}\delta_{k-1m} + d_{ijkk}\delta_{0p}\delta_{0q}\delta_{km-1}, 0). \end{cases}$ | |

Table 3 Second derivatives of $x$, $y$, and $z$ with respect to time

| | |
|---|---|
| $\omega^2 \dfrac{\partial^2 x}{\partial \theta_1^2} \to \begin{cases} -\omega_0^2 p^2(x_{ijkm}^{pq}, \bar{x}_{ijkm}^{pq}) \\ (-\Omega_{i-1jkk}\delta_{1p}\delta_{0q}\delta_{km}, 0), \end{cases}$ | $2\omega v \dfrac{\partial^2 x}{\partial \theta_1 \partial \theta_2} \to -2\omega_0 v_0 pq(x_{ijkm}^{pq}, \bar{x}_{ijkm}^{pq})$ |
| $v^2 \dfrac{\partial^2 x}{\partial \theta_2^2} \to -v_0^2 q^2(x_{ijkm}^{pq}, \bar{x}_{ijkm}^{pq}),$ | $2\omega\lambda \dfrac{\partial^2 x}{\partial \theta_1 \partial \theta_3} \to 2\omega_0 \lambda_0 p(k-m)(\bar{x}_{ijkm}^{pq}, -x_{ijkm}^{pq})$ |
| $\lambda^2 \dfrac{\partial^2 x}{\partial \theta_3^2} \to \begin{cases} \lambda_0^2(k-m)^2(x_{ijkm}^{pq}, \bar{x}_{ijkm}^{pq}) \\ (\Lambda_{ijmm}\delta_{0p}\delta_{0q}\delta_{k-1m} - \Lambda_{ijkk}\delta_{0p}\delta_{0q}\delta_{km-1}, 0), \end{cases}$ | $2v\lambda \dfrac{\partial^2 x}{\partial \theta_2 \partial \theta_3} \to 2v_0 \lambda_0 q(k-m)(\bar{x}_{ijkm}^{pq}, -x_{ijkm}^{pq})$ |
| $\omega^2 \dfrac{\partial^2 y}{\partial \theta_1^2} \to \begin{cases} -\omega_0^2 p^2(y_{ijkm}^{pq}, \bar{y}_{ijkm}^{pq}) \\ (0, -\Omega_{i-1jkk}\kappa_1\delta_{1p}\delta_{0q}\delta_{km}), \end{cases}$ | $2\omega v \dfrac{\partial^2 y}{\partial \theta_1 \partial \theta_2} \to -2\omega_0 v_0 pq(y_{ijkm}^{pq}, \bar{y}_{ijkm}^{pq})$ |
| $v^2 \dfrac{\partial^2 y}{\partial \theta_2^2} \to -v_0^2 q^2(y_{ijkm}^{pq}, \bar{y}_{ijkm}^{pq}),$ | $2\omega\lambda \dfrac{\partial^2 y}{\partial \theta_1 \partial \theta_3} \to 2\omega_0 \lambda_0 p(k-m)(\bar{y}_{ijkm}^{pq}, -y_{ijkm}^{pq})$ |
| $\lambda^2 \dfrac{\partial^2 y}{\partial \theta_3^2} \to \begin{cases} \lambda_0^2(k-m)^2(y_{ijkm}^{pq}, \bar{y}_{ijkm}^{pq}) \\ (\kappa_2\Lambda_{ijmm}\delta_{0p}\delta_{0q}\delta_{k-1m} - \kappa_2\Lambda_{ijkk}\delta_{0p}\delta_{0q}\delta_{km-1}, 0), \end{cases}$ | $2v\lambda \dfrac{\partial^2 y}{\partial \theta_2 \partial \theta_3} \to 2v_0 \lambda_0 q(k-m)(\bar{y}_{ijkm}^{pq}, -y_{ijkm}^{pq})$ |
| $\omega^2 \dfrac{\partial^2 z}{\partial \theta_1^2} \to \begin{cases} -\omega_0^2 p^2(z_{ijkm}^{pq}, \bar{z}_{ijkm}^{pq}) \\ (-\Omega_{i-1jkk}\eta\delta_{1p}\delta_{0q}\delta_{km}, 0) \end{cases}$ | $2\omega v \dfrac{\partial^2 z}{\partial \theta_1 \partial \theta_2} \to -2\omega_0 v_0 pq(z_{ijkm}^{pq}, \bar{z}_{ijkm}^{pq})$ |
| $v^2 \dfrac{\partial^2 z}{\partial \theta_2^2} \to \begin{cases} -v_0^2 q^2(z_{ijkm}^{pq}, \bar{z}_{ijkm}^{pq}) \\ (-N_{ij-1kk}\delta_{0p}\delta_{1q}\delta_{km}, 0) \end{cases}$ | $2\omega\lambda \dfrac{\partial^2 z}{\partial \theta_1 \partial \theta_3} \to 2\omega_0 \lambda_0 p(k-m)(\bar{z}_{ijkm}^{pq}, -z_{ijkm}^{pq})$ |
| $\lambda^2 \dfrac{\partial^2 z}{\partial \theta_3^2} \to \begin{cases} \lambda_0^2(k-m)^2(z_{ijkm}^{pq}, \bar{z}_{ijkm}^{pq}) \\ (\kappa_3\eta\Lambda_{ijmm}\delta_{0p}\delta_{0q}\delta_{k-1m} - \kappa_3\eta\Lambda_{ijkk}\delta_{0p}\delta_{0q}\delta_{km-1}, 0) \end{cases}$ | $2v\lambda \dfrac{\partial^2 z}{\partial \theta_2 \partial \theta_3} \to 2v_0 \lambda_0 q(k-m)(\bar{z}_{ijkm}^{pq}, -z_{ijkm}^{pq})$ |

Following this, we relocate all the known terms to the right side of (13), add them to ($r_{ijkm}^{pq}, \bar{r}_{ijkm}^{pq}$), ($s_{ijkm}^{pq}, \bar{s}_{ijkm}^{pq}$) and ($t_{ijkm}^{pq}, \bar{t}_{ijkm}^{pq}$), which are re-denoted with the same name for clarity. In summary, the linear equations of *n*-order unknown coefficients are obtained by identifying the *n*-order terms in both sides of (13). Depending on the specific situations, these linear equations exhibit slight variations as follows:

**Case 1**: $p = 1$, $q = 0$, $k = m$.

$$\begin{bmatrix} -\omega_0^2 + \gamma_1 & -2\omega_0 & 0 & 0 \\ -2\omega_0 & -\omega_0^2 + \gamma_2 & 0 & 0 \\ 0 & 0 & -\omega_0^2 + \gamma_1 & 2\omega_0 \\ 0 & 0 & 2\omega_0 & -\omega_0^2 + \gamma_2 \end{bmatrix} \begin{bmatrix} x_{ijkm}^{pq} \\ \overline{y}_{ijkm}^{pq} \\ \overline{x}_{ijkm}^{pq} \\ y_{ijkm}^{pq} \end{bmatrix} + \begin{bmatrix} -2(\kappa_1 + \omega_0)\omega_{i-1jkk} \\ -2(1+\kappa_1\omega_0)\omega_{i-1jkk} \\ 0 \\ 0 \end{bmatrix} = \begin{bmatrix} r_{ijkm}^{pq} + \overline{\Omega}_{i-1jkk} \\ \overline{s}_{ijkm}^{pq} + \kappa_1\overline{\Omega}_{i-1jkk} \\ \overline{r}_{ijkm}^{pq} \\ s_{ijkm}^{pq} \end{bmatrix}$$

$$\begin{bmatrix} -\omega_0^2 + c_2 & 0 \\ 0 & -\omega_0^2 + c_2 \end{bmatrix} \begin{bmatrix} z_{ijkm}^{pq} \\ \overline{z}_{ijkm}^{pq} \end{bmatrix} + \begin{bmatrix} d_{i-1jkk} \\ 0 \end{bmatrix} = \begin{bmatrix} t_{ijkm}^{pq} \\ \overline{t}_{ijkm}^{pq} \end{bmatrix}$$ (20)

where $\gamma_1 = -1 - 2c_2$, $\gamma_2 = c_2 - 1$. In this case, $\overline{r}_{ijkm}^{pq} = \overline{s}_{ijkm}^{pq} = \overline{t}_{ijkm}^{pq} = 0$. Thus $\overline{x}_{ijkm}^{pq} = \overline{y}_{ijkm}^{pq} = \overline{z}_{ijkm}^{pq} = 0$, and (20) becomes

$$\left(-\omega_0^2 + \gamma_1\right) x_{ijkm}^{pq} - 2\omega_{00} \overline{y}_{ijkm}^{pq} - 2(\omega_{00} + \kappa)\omega_{i-1jkk} = r_{ijkm}^{pq} + \overline{\Omega}_{i-1jkk}$$

$$-2\omega_0 x_{ijkm}^{pq} + (c_2 - 1 - \omega_{00}^2) \overline{y}_{ijkm}^{pq} - 2(\kappa\omega_{00} + 1)\omega_{i-1jkk} = \overline{s}_{ijkm}^{pq} + \kappa_1 \overline{\Omega}_{i-1jkk}$$

$$\left(-\omega_0^2 + c_2\right) z_{ijkm}^{pq} - d_{i-1jkk} = t_{ijkm}^{pq}.$$

Due to the definition of $\omega_0$, $x_{ijkm}$ and $y_{ijkm}$ are not independent. Letting $x_{ijkm} = 0$, $y_{ijkm}$ and $\omega_{i-1jkk}$ can be solved. For the third equation, set $z_{ijkm} = 0$, and then $d_{i-1jkk} = -t_{ijkm}^{pq}$ to ensure the coefficient of coupling factor is not zero.

**Case 2**: $p = 0$, $q = 1$, $k = m$.

$$\begin{bmatrix} -v_0^2 + \gamma_1 & -2v_0 & 0 & 0 \\ -2v_0 & -v_0^2 + \gamma_2 & 0 & 0 \\ 0 & 0 & -v_0^2 + \gamma_1 & 2v_0 \\ 0 & 0 & 2v_0 & -v_0^2 + \gamma_2 \end{bmatrix} \begin{bmatrix} x_{ijkm}^{pq} \\ \overline{y}_{ijkm}^{pq} \\ \overline{x}_{ijkm}^{pq} \\ y_{ijkm}^{pq} \end{bmatrix} = \begin{bmatrix} r_{ijkm}^{pq} \\ \overline{s}_{ijkm}^{pq} \\ \overline{r}_{ijkm}^{pq} \\ s_{ijkm}^{pq} \end{bmatrix}$$

$$\begin{bmatrix} -v_0^2 + c_2 & 0 \\ 0 & -v_0^2 + c_2 \end{bmatrix} \begin{bmatrix} z_{ijkm}^{pq} \\ \overline{z}_{ijkm}^{pq} \end{bmatrix} + \begin{bmatrix} -2v_0 v_{ij-1km} \\ 0 \end{bmatrix} = \begin{bmatrix} t_{ijkm}^{pq} + \overline{N}_{ij-1km} \\ \overline{t}_{ijkm}^{pq} \end{bmatrix}$$ (21)

In this case, the paired coefficients ($x_{ijkm}^{pq}, \overline{x}_{ijkm}^{pq}$), ($y_{ijkm}^{pq}, \overline{y}_{ijkm}^{pq}$) can be solved directly. From $\overline{s}_{ijkm}^{pq} = 0$ we have $\overline{z}_{ijkm}^{pq} = 0$. Due to the definition of $v_0$, $-v_0^2 + c_2 = 0$, and thus $z_{ijkm}^{pq}$ is set zero, $-t_{ijkm}^{pq}/2v_0$.

**Case 3**: $p = q = 0$, $k - m = 1$.

$$\begin{bmatrix} \lambda_0^2 + \gamma_1 & 0 & 0 & -2\lambda_0 \\ 0 & \lambda_0^2 + \gamma_2 & 2\lambda_0 & 0 \\ 0 & -2\lambda_0 & \lambda_0^2 + \gamma_1 & 0 \\ 2\lambda_0 & 0 & 0 & \lambda_0^2 + \gamma_2 \end{bmatrix} \begin{bmatrix} x_{ijkm}^{pq} \\ \overline{y}_{ijkm}^{pq} \\ \overline{x}_{ijkm}^{pq} \\ y_{ijkm}^{pq} \end{bmatrix} + \begin{bmatrix} 2(\lambda_0 - \kappa_2)\lambda_{ijmm} \\ 0 \\ 0 \\ 2(\kappa_2\lambda_0 + 1)\lambda_{ijmm} \end{bmatrix} = \begin{bmatrix} r_{ijkm}^{pq} - \overline{\Lambda}_{ijmm} \\ \overline{s}_{ijkm}^{pq} \\ \overline{r}_{ijkm}^{pq} \\ s_{ijkm}^{pq} - \kappa_2\overline{\Lambda}_{ijmm} \end{bmatrix}$$

$$\begin{bmatrix} \lambda_0^2 + c_2 & 0 \\ 0 & \lambda_0^2 + c_2 \end{bmatrix} \begin{bmatrix} z_{ijkm}^{pq} \\ \overline{z}_{ijkm}^{pq} \end{bmatrix} + \begin{bmatrix} d_{ijmm} \\ 0 \end{bmatrix} = \begin{bmatrix} t_{ijkm}^{pq} \\ \overline{t}_{ijkm}^{pq} \end{bmatrix}$$ (22)

In this case, $\bar{r}^{pq}_{ijkm}=\bar{s}^{pq}_{ijkm}=\bar{t}^{pq}_{ijkm}=0$. Thus $\bar{x}^{pq}_{ijkm}=\bar{y}^{pq}_{ijkm}=\bar{z}^{pq}_{ijkm}=0$, and (22) becomes

$$\left(\lambda_0^2+\gamma_1\right)x^{pq}_{ijkm}-2\lambda_0 y^{pq}_{ijkm}+2(\lambda_0-\kappa_2)\lambda_{ijmm}=r^{pq}_{ijkm}-\bar{\Lambda}_{ijmm}$$

$$2\lambda_0 x^{pq}_{ijkm}+\left(\lambda_0^2+\gamma_2\right)y^{pq}_{ijkm}+2(\kappa_2\lambda_0+1)\lambda_{ijmm}=s^{pq}_{ijkm}-\kappa_2\bar{\Lambda}_{ijmm}$$

$$\left(\lambda_0^2+c_2\right)z^{pq}_{ijkm}+d_{ijmm}=t^{pq}_{ijkm}$$

Similar to Case 1, $x^{pq}_{ijkm}$ and $y^{pq}_{ijkm}$ are are not independent. Letting $x^{pq}_{ijkm}=0$, $y^{pq}_{ijkm}$ and $\lambda_{i\text{-}1jkk}$ can be solved. For the third equation, $d_{ijmm}$ is given in Case 1, and then $z^{pq}_{ijkm}=\left(t^{pq}_{ijkm}-d_{ijmm}\right)/\left(\lambda_0^2+c_2\right)$.

**Case 4**: $p=q=0$, $k-m=-1$.

$$\begin{bmatrix} \lambda_0^2+\gamma_1 & 0 & 0 & 2\lambda_0 \\ 0 & \lambda_0^2+\gamma_2 & -2\lambda_0 & 0 \\ 0 & 2\lambda_0 & \lambda_0^2+\gamma_1 & 0 \\ -2\lambda_0 & 0 & 0 & \lambda_0^2+\gamma_2 \end{bmatrix}\begin{bmatrix} x^{pq}_{ijkm} \\ \bar{y}^{pq}_{ijkm} \\ \bar{x}^{pq}_{ijkm} \\ y^{pq}_{ijkm} \end{bmatrix}+\begin{bmatrix} 2(\lambda_0-\kappa_2)\lambda_{ijkk} \\ 0 \\ 0 \\ 2(\kappa_2\lambda_0+1)\lambda_{ijkk} \end{bmatrix}=\begin{bmatrix} r^{pq}_{ijkm}-\Lambda_{ijkk} \\ \bar{s}^{pq}_{ijkm} \\ \bar{r}^{pq}_{ijkm} \\ s^{pq}_{ijkm}+\kappa_2\Lambda_{ijkk} \end{bmatrix} \quad (23)$$

$$\begin{bmatrix} \lambda_0^2+c_2 & 0 \\ 0 & \lambda_0^2+c_2 \end{bmatrix}\begin{bmatrix} z^{pq}_{ijkm} \\ \bar{z}^{pq}_{ijkm} \end{bmatrix}+\begin{bmatrix} d_{ijkk} \\ 0 \end{bmatrix}=\begin{bmatrix} t^{pq}_{ijkm} \\ \bar{t}^{pq}_{ijkm} \end{bmatrix}$$

This case is similar to Case 3. We have $x^{pq}_{ijkm}=\bar{x}^{pq}_{ijkm}=\bar{y}^{pq}_{ijkm}=\bar{z}^{pq}_{ijkm}=0$, $y^{pq}_{ijkm}$ and $\lambda_{i\text{-}1jkk}$ is solved by the following equation

$$-2\lambda_0 y^{pq}_{ijkm}+2(\lambda_0-\kappa_2)\lambda_{ijkk}=r^{pq}_{ijkm}-\Lambda_{ijkk}$$

$$\left(\lambda_0^2+\gamma_2\right)y^{pq}_{ijkm}+2(\kappa_2\lambda_0+1)\lambda_{ijkk}=s^{pq}_{ijkm}+\kappa_2\Lambda_{ijkk}$$

$z^{pq}_{ijkm}$ is given by $z^{pq}_{ijkm}=\left(t^{pq}_{ijkm}-d_{ijkk}\right)/\left(\lambda_0^2+c_2\right)$.

**Otherwise**: All the remaining cases.

$$\begin{bmatrix} \Psi^2-1-2c_2 & -2\varpi_{pq} & 2\lambda_0(k-m)\varpi_{pq} & -2\lambda_0(k-m) \\ -2\varpi_{pq} & \Psi^2+c_2-1 & 2\lambda_0(k-m) & -2\lambda_0(k-m)\varpi_{pq} \\ -2\lambda_0(k-m)\varpi_{pq} & -2\lambda_0(k-m) & \Psi^2-1-2c_2 & 2\varpi_{pq} \\ 2\lambda_0(k-m) & 2\lambda_0(k-m)\varpi_{pq} & 2\varpi_{pq} & \Psi^2+c_2-1 \end{bmatrix}\begin{bmatrix} x^{pq}_{ijkm} \\ \bar{y}^{pq}_{ijkm} \\ \bar{x}^{pq}_{ijkm} \\ y^{pq}_{ijkm} \end{bmatrix}=\begin{bmatrix} r^{pq}_{ijkm} \\ \bar{s}^{pq}_{ijkm} \\ \bar{r}^{pq}_{ijkm} \\ s^{pq}_{ijkm} \end{bmatrix} \quad (24)$$

$$\begin{bmatrix} \Psi^2+c_2 & 2\lambda_0(k-m)\varpi_{pq} \\ -2\lambda_0(k-m)\varpi_{pq} & \Psi^2+c_2 \end{bmatrix}\begin{bmatrix} z^{pq}_{ijkm} \\ \bar{z}^{pq}_{ijkm} \end{bmatrix}=\begin{bmatrix} t^{pq}_{ijkm} \\ \bar{t}^{pq}_{ijkm} \end{bmatrix}$$

where $\varpi_{pq}=p\omega_0+qv_0$, $\Psi^2=\lambda_0^2(k-m)^2-\varpi_{pq}^2$. The determinants of coefficient matrix of (24) are always non-zero. The paired coefficients ($x^{pq}_{ijkm}, \bar{x}^{pq}_{ijkm}$), ($y^{pq}_{ijkm}, \bar{y}^{pq}_{ijkm}$), ($z^{pq}_{ijkm}, \bar{z}^{pq}_{ijkm}$) can be solved directly.

## 4. Results and discussions

In this section, the analytical bifurcation equation is derived during the computation of the third-order solution for phase space near collinear libration points in the RTBP with arbitrary system parameter $\mu$. Furthermore, the construction of the series solution up to certain $n$ order is implemented for the given system parameter $\mu$, such as the Sun-Earth system ($\mu$ = 3.040423398444176e-6) or Earth-Moon system ($\mu = 1.215058191870689e-2$) using the C++ 17 programming language.

**4.1 Bifurcation analysis**

It is noted that $\Delta = 0$ exists solution $\eta(\alpha_1, \alpha_2, \alpha_3, \alpha_4)$ only when the series solution (16) is of the third order or higher. Thus, the bifurcation analysis of various orbits in phase space near collinear libration points begins with the third-order bifurcation equation $\Delta = 0$. Following the analytical construction method outlined in Section 3, the third-order bifurcation equation is given as:

$$\begin{aligned}\Delta &= d_{0000} + d_{2000}\alpha_1^2 + d_{0200}\alpha_2^2 + d_{0011}\alpha_3\alpha_4 \\ &= \left(l_1\alpha_1^2 + l_2\alpha_3\alpha_4\right)\eta^4 + \left(l_3\alpha_1^2 + l_4\alpha_2^2 + l_5\alpha_3\alpha_4\right)\eta^2 + l_6\alpha_1^2 + l_7\alpha_2^2 + l_8\alpha_3\alpha_4 - (\omega_0^2 - v_0^2) \\ &= a\eta^4 + b\eta^2 + c = 0\end{aligned} \quad (25)$$

where $a = l_1\alpha_1^2 + l_2\alpha_3\alpha_4, b = l_3\alpha_1^2 + l_4\alpha_2^2 + l_5\alpha_3\alpha_4, c = l_6\alpha_1^2 + l_7\alpha_2^2 + l_8\alpha_3\alpha_4 - (\omega_0^2 - v_0^2)$, $l_i$ ($i = 1, 2, \ldots, 8$) are constant related to the system parameter $\mu$. Their variations with the parameter $\mu$ are illustrated in Fig. 1. It can be observed that in $a$, $b$, and $c$, the coefficient of $\alpha_1^2$ is significantly larger than the coefficients of $\alpha_2^2$ and $\alpha_3\alpha_4$. Therefore, the bifurcation induced by the variation in the amplitude of $\alpha_1$ is the most pronounced and may overshadow bifurcations caused by changes in the other amplitudes. This is also why we initially observe the bifurcation of halo orbits from planar Lyapunov orbits. However, Equation (25) implies that bifurcations of other types of orbits also exist. The following detailed analysis will be conducted on them.

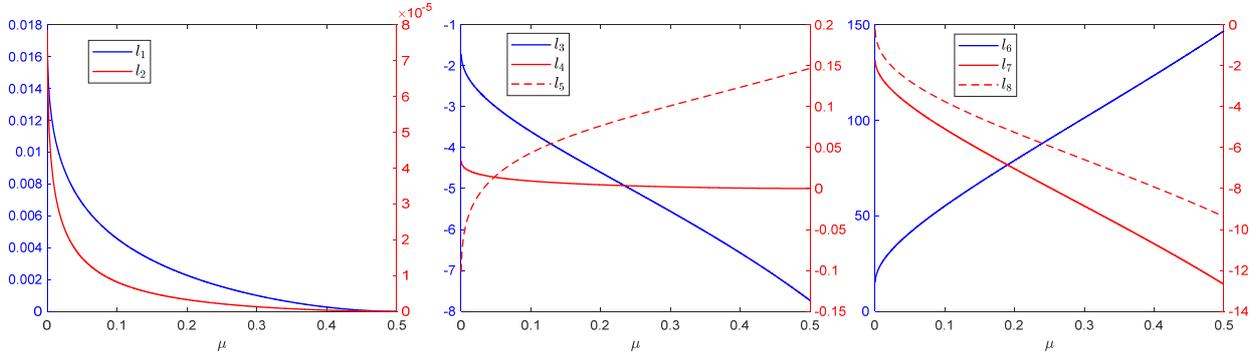

Fig. 1 Coefficients of third-order bifurcation equation for $L_1$

Equation (25) is quadratic equation in terms of $\eta^2$, establishing the relationship between $\eta$ and $\alpha_i$ ($i = 1,2,3,4$). When the amplitudes are all zero, (25) have obviously no solution. As the amplitudes $\alpha_i$ vary, the changes in the coefficients $a$, $b$, and $c$ become complicated, resulting in (25) having various types of real solutions and different critical bifurcation conditions. Next, we discuss the three cases for critical bifurcation conditions with reasonable values of $\alpha_i$:

**Case 1**. $D > 0$, $-b/a > 0$, $c = 0$, where $D = b^2 - 4ac$.

In this case, we have

$$c = l_6\alpha_1^2 + l_7\alpha_2^2 + l_8\alpha_3\alpha_4 - \left(\omega_0^2 - v_0^2\right) = 0, \quad (26)$$

Figure 1 shows that $l_6 > 0$, $l_7 < 0$, and $l_8 < 0$ for the three collinear libration points with all system parameter $\mu \in (0, 0.5]$. Thus, (26) is a hyperboloid about $\alpha_1^2$, $\alpha_2^2$, and $\alpha_3\alpha_4$, depicted by the blue surface in Fig. 1(a). On this bifurcation surface there are a trivial solution $\eta = 0$ and a non-trivial solution $\eta^2 = -\frac{b}{a}$. On the right side of the hyperboloid, (26) has two solutions:

$$\eta^2 = \frac{-b \pm \sqrt{D}}{2a} > 0 \tag{27}$$

Thus, there are total four feasible solutions for $\eta = \pm\sqrt{\frac{-b \pm \sqrt{D}}{2a}}$. However, on the left side of the hyperboloid, (26) has only one positive real solution. Thus, there are two feasible solutions for $\eta = \pm\sqrt{\frac{-b + \sqrt{D}}{2a}}$. When considering only the central manifold component and their manifolds, i.e., $\alpha_3\alpha_4 = 0$, the hyperboloid bifurcation equation degenerates into a hyperbolic bifurcation equation, as shown in the red curve in Fig. 2(a). Practically, if $\alpha_2 = \alpha_3 = \alpha_4 = 0$, it will further degenerate into a bifurcation point of planer Lyapunov orbits. It is denoted by a black pentagram in Fig. 2(a), where a classical halo orbit is generated.

**Case 2**. $D > 0$, $a = 0$, $bc < 0$

In this case, we have

$$a = l_1\alpha_1^2 + l_2\alpha_3\alpha_4 = 0, \tag{28}$$

It is a paraboloid about $\alpha_1^2$, and $\alpha_3\alpha_4$, depicted by the green surface in Fig. 1(a). $\eta = \pm\sqrt{\frac{-b}{c}}$ on this bifurcation surface. On the right side of the paraboloid, $a > 0$, $-b/2a < 0$, and (28) has two feasible solutions for $\eta = \pm\sqrt{\frac{-b + \sqrt{D}}{2a}}$. On the left side, (28) has four feasible solutions for $\eta = \pm\sqrt{\frac{-b \pm \sqrt{D}}{2a}}$. This critical bifurcation condition occurs only when both $\alpha_3$ and $\alpha_4$ are non-zero, and it cannot be found in the case of traditional central manifold and their invariant manifolds.

**Case 3**. $D = 0$, $-b/2a > 0$,

In this case, we have

$$\left(l_3\alpha_1^2 + l_4\alpha_2^2 + l_5\alpha_3\alpha_4\right)^2 - 4\left(l_1\alpha_1^2 + l_2\alpha_3\alpha_4\right)\left(l_6\alpha_1^2 + l_7\alpha_2^2 + l_8\alpha_3\alpha_4 - (\omega_0^2 - v_0^2)\right) = 0 \tag{29}$$

Equation (29) describes a complicated quartic surface about $\alpha_1^2$, $\alpha_2^2$, and $\alpha_3\alpha_4$ as shown by the red surfaces in Fig. 2(a). $\eta = \pm\sqrt{\frac{-b}{2a}}$ on these bifurcation surfaces. There is no solution inside the surfaces, while there are four solutions for $\eta = \frac{-b \pm \sqrt{D}}{2a}$ outside the surfaces. Figure 2(b) illustrates the distribution of the number of solutions for the coupling coefficient $\eta$ in the interval $\alpha_1 \in [0, 0.5]$, $\alpha_2 \in [0, 1.0]$, and $\alpha_3\alpha_4 \in [-0.5, 0.5]$ in the Earth-Sun system. This numerical result is consistent with the analytical findings discussed above.

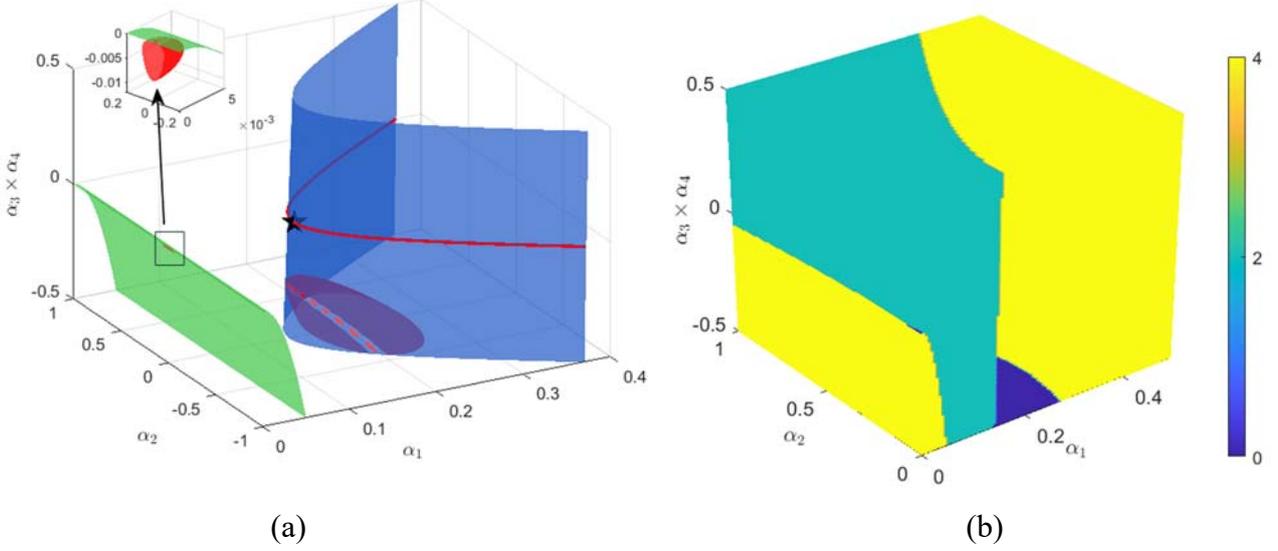

(a)           (b)

Fig. 2 The distribution of the third-order bifurcation solutions. (a) bifurcation surfaces in three different cases. (b) the distribution of the number of solutions for the coupling coefficient $\eta$ in the interval $\alpha_1 \in [0, 0.5]$, $\alpha_2 \in [0, 1.0]$, and $\alpha_3\alpha_4 \in [-0.5, 0.5]$.

*Remark 3*. Third-order bifurcation equation $\Delta = 0$ implies that the coupling coefficients depend solely on three parameters: $\alpha_1^2$, $\alpha_2^2$, and $\alpha_3\alpha_4$. This remains valid when considering solutions of higher orders. This suggests that the number of parameters in the bifurcation equation is determined solely by the system's degrees of freedom. For halo orbits, which consider only one parameter $\alpha_1^2$, the critical conditions for bifurcation correspond to a single point in a one-parameter space. For quasihalo orbits, considering two parameters $\alpha_1^2$ and $\alpha_2^2$, the bifurcation critical conditions correspond to a one-dimensional curve in a two-parameter plane. When considering all parameters, the bifurcation critical conditions correspond to a two-dimensional surface in a three-parameter space. Naturally, one can infer that when considering Hamiltonian systems with higher degrees of freedom ($N > 3$), the critical conditions for bifurcation will generally correspond to an $N - 1$ dimensional hypersurface in $N$-dimensional space.

*Remark 4*. We can see from Case 1 that the solution $\eta = \pm\sqrt{\dfrac{-b-\sqrt{D}}{2a}}$ from $\Delta = 0$ corresponds to two families of classical halo orbit bifurcating from planar Lyapunov orbits when $\alpha_1$ satisfies $c = l_6\alpha_1^2 - (\omega_0^2 - v_0^2) \geq 0$. However, $\Delta = 0$ has another solution $\eta = \pm\sqrt{\dfrac{-b+\sqrt{D}}{2a}}$ whether $\alpha_1$ satisfies $c \geq 0$ or $c < 0$, which means there may exists another two families of periodic orbit. These two families of orbits do not bifurcate at a certain amplitude of planar Lyapunov orbits, but rather coexist with planar Lyapunov orbits. Due to the similarity in the generation of $\eta$ value to halo orbits, they can be referred to as the second type of halo orbits.

### 4.2 Numerical results

In this subsection, the semi-analytical solution of phase space near collinear libration points for the given system parameter is computed up to order 23. The accuracy of various types of orbits around collinear libration points is analyzed.

**Case $\alpha_3 = \alpha_4 = 0$**: Center manifolds

As mentioned in Section 3, the solution (16) describes the center manifold portion in the CRTBP when $\alpha_3 = \alpha_4 = 0$. If the values of $\alpha_1$ and $\alpha_2$ are small, $\Delta = 0$ has one pair of opposite real solutions. In this case, if $\eta = 0$, (16) describes planar Lyapunov orbits, vertical Lyapunov orbits, and Lissajous orbits as shown in Fig. 3(a). If $\eta \neq 0$, (16) describes two new families (second type) of halo orbits and their corresponding quasihalo orbits as shown in Fig. 3(b). These orbits are similar to classical halo/quasihalo orbits, but with larger $\eta$ values, exhibiting high levels of instability. With the increment in $\alpha_1$ and $\alpha_2$, $\Delta = 0$ will have additional pairs of real solutions. This leads to the bifurcation of not only classical halo/quasihalo orbits but also previously new families of quasihalo orbits emerging from planar Lyapunov/Lissajous orbit. Fig. 3(c) and (d) represent two quasihalo orbits with identical amplitudes but different coupling coefficients. The former has a small coupling coefficient, resembling Lissajous orbits more closely, while the latter, with a larger coupling coefficient, exhibits characteristics typical of classical quasihalo orbits.

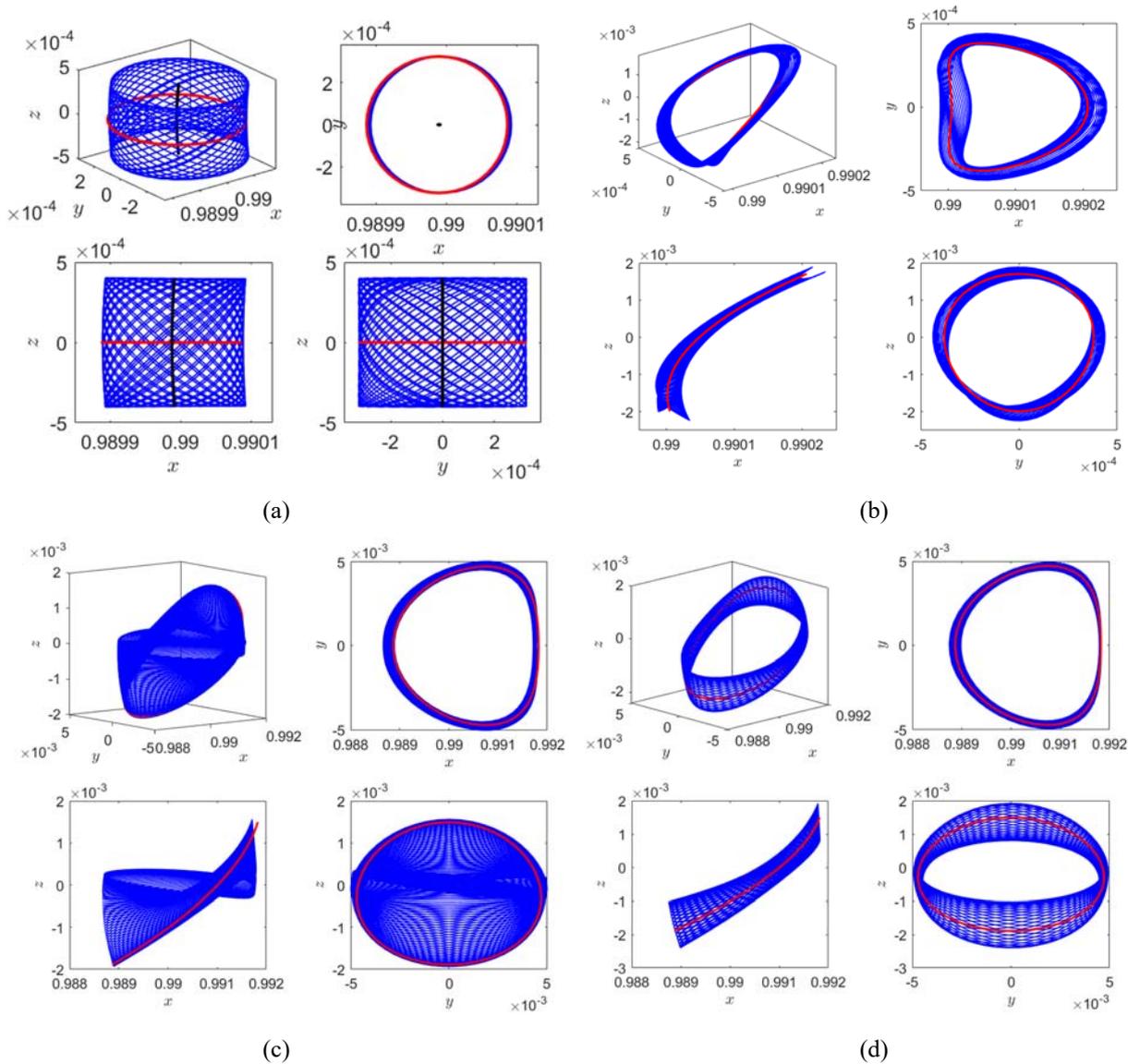

(a) (b)

(c) (d)

Fig. 3 Periodic/quasi-periodic orbits in the center manifolds and their projection on *xy*, *xz*, and *yz* planes. (a) Red: planar Lyapunov orbit; black: vertical Lyapunov orbit; blue: Lissajous orbit with $\alpha_1 = 0.01$, $\alpha_2 = 0.04$. (b) Second-type halo orbit and quasihalo orbit with $\alpha_1 = 0.01$, $\alpha_2 = 0.04$. (c) and (d) Halo orbit, quasihalo orbits with $\alpha_1 = 0.16$, $\alpha_2 = 0.04$.

**Case $\alpha_1 = \alpha_2 = 0$**: Hyperbolic manifolds of collinear libration points

The solution (16) describes hyperbolic manifolds portion of collinear libration point in the CRTBP when $\alpha_1 = \alpha_2 = 0$. Based on the analysis in Subsection 4.1, we known that $\Delta = 0$ has a non-zero solution when $\alpha_3 \times \alpha_4 \neq 0$. Figure 4 illustrates that when $\alpha_3 \times \alpha_4 > 0$, there exists a pair of real solutions with relatively large value of $|\eta|$. In the case of $\alpha_3 \times \alpha_4 < 0$, two pairs of real solutions emerge beyond a certain critical value, characterized by smaller values of $|\eta|$, indicating a lesser degree of coupling compared to the case of $\alpha_3 \times \alpha_4 > 0$. Figures 5 shows the distribution of orbits when the coupling coefficient is both zero and non-zero. When $\eta = 0$, all the orbits with $\alpha_3 \times \alpha_4 \neq 0$ is restricted in the *xy* plane and divided into four parts by stable/unstable manifolds of collinear libration point as shown in Fig. 5(a). When $\eta \neq 0$, new orbits bifurcate, extending beyond the *xy*-plane. Additionally, with a larger coupling coefficient, the amplitude of motion in the *z*-direction increases.

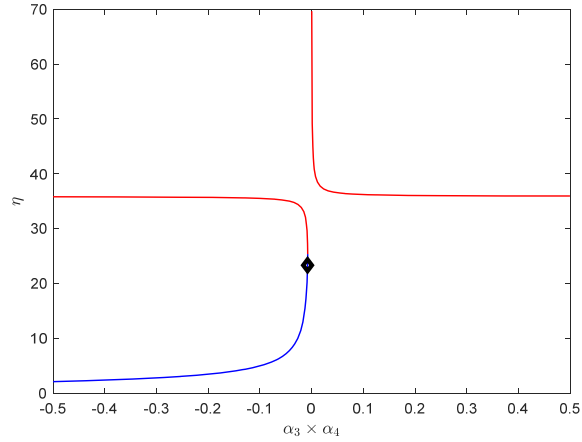

Fig. 4 The variation of $\eta$ with $\alpha_3 \times \alpha_4$ in the case $\alpha_1 = \alpha_2 = 0$.

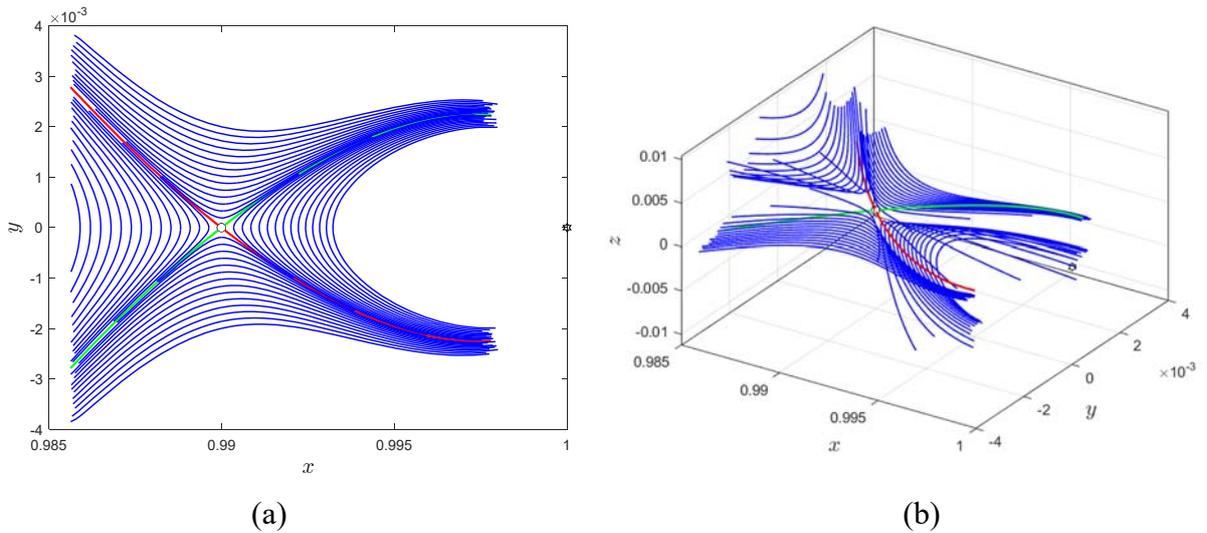

(a)          (b)

Fig. 5 Hyperbolic subspace. Green: stable manifolds. Red: unstable manifolds. Blue: (a) Orbits with $\alpha_3 \times \alpha_4 \neq 0$ restricted in the *xy* plane. (b) Coupling orbits with $\alpha_3 \times \alpha_4 \neq 0$.

**Case $α_3 ≠ 0$, $α_4 = 0$ ($α_3 = 0$, $α_4 ≠ 0$)**: Hyperbolic manifolds of periodic/quasi-periodic orbit

The solution (16) describes unstable (stable) manifolds of periodic/quasi-periodic orbits in the CRTBP when $α_3 ≠ 0$, $α_4 = 0$ ($α_3 = 0$, $α_4 ≠ 0$). By selecting two set of parameters $α_3 > 0$, $α_4 = 0$, and $α_3 < 0$, $α_4 = 0$, along with a positive time interval, we can obtain two branches of unstable manifolds of Lissajous orbits ($α_1 ≠ 0$, $α_2 ≠ 0$, $η = 0$), quasihalo orbits ($α_1 ≠ 0$, $α_2 ≠ 0$, $η ≠ 0$), halo orbits ($α_1 ≠ 0$, $α_2 = 0$, $η ≠ 0$), and the second-type of halo orbits ($α_1 ≠ 0$, $α_2 = 0$, $η ≠ 0$), as shown by the blue curves in Fig. 6. Similar, by selecting $α_3 = 0$, $α_4 > 0$, and $α_3 = 0$, $α_4 < 0$, along with a negative time interval, two branches of stable manifolds of these periodic/quasi-periodic orbits are obtained as shown by the red curves in Fig. 6.

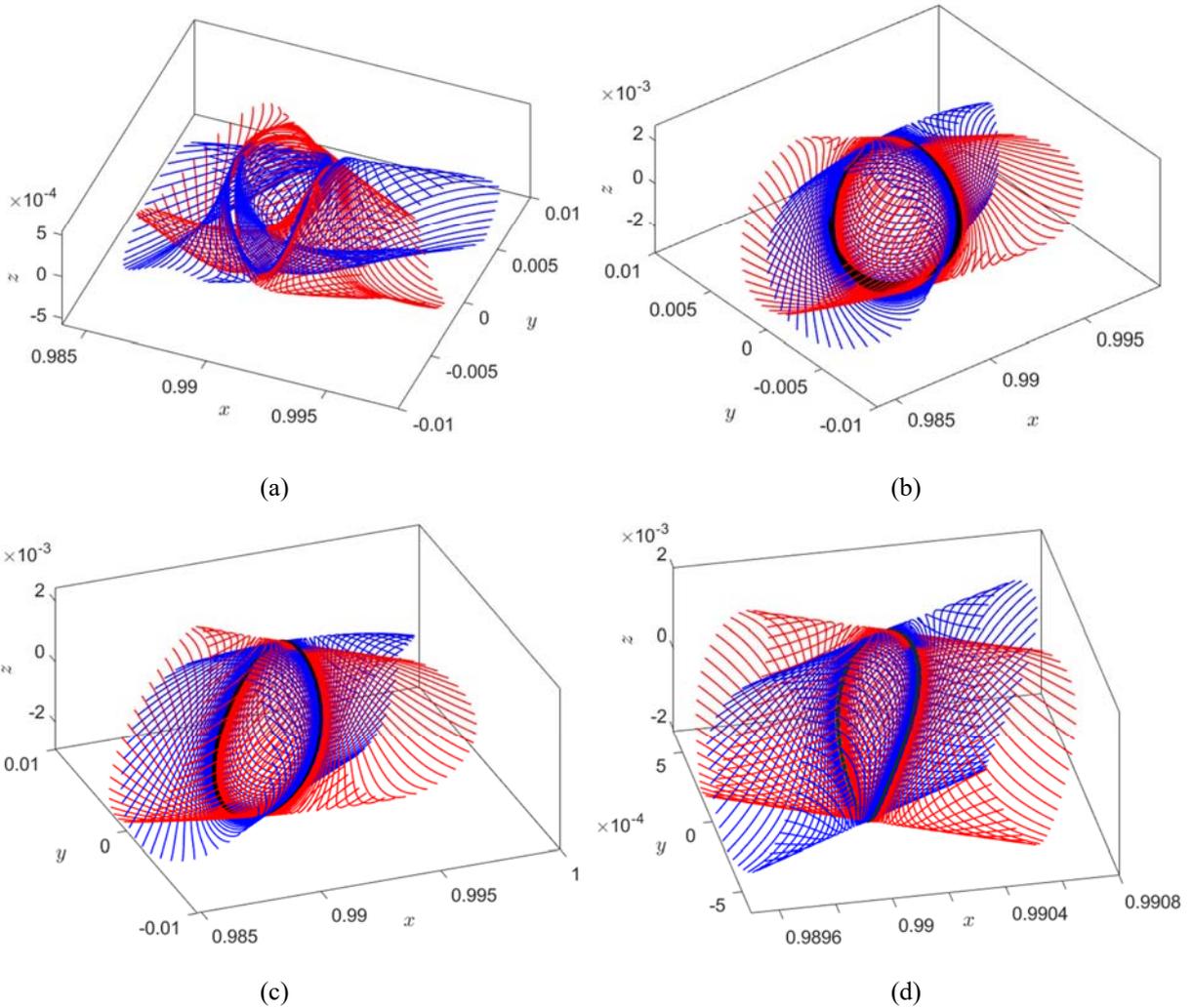

Fig. 6 Periodic/quasi-periodic orbits and their invariant manifolds. Black: Periodic/quasi-periodic orbits. Blue: unstable manifolds. Red: stable manifolds. (a) Lissajous orbit with $α_1 = 0.16$, $α_2 = 0.02$, $η = 0$. (b) Quasihalo orbit with $α_1 = 0.16$, $α_2 = 0.02$, $η = 1.4556115$. (c) Halo orbit with $α_1 = 0.16$, $α_2 = 0$, $η = 1.4686092$. (d) Second-type halo orbit with $α_1 = 0.01$, $α_2 = 0$, $η = 18.8704922$.

**Case $α_3 ≠ 0$, $α_4 ≠ 0$**: Transit and non-transit orbits

The solution (16) describes transit and non-transit orbits of periodic/quasi-periodic orbits in the CRTBP when $α_3α_4 < 0$ ($α_3α_4 > 0$). Figure 7(a) and (b) presents the transit orbits of halo orbits and quasihalo orbits with $α_3 < 0$ and $α_4 > 0$. In position space, these orbits traverse from the left side of the halo/quasihalo orbit to the right side of the halo/quasihalo orbit. If $α_3 > 0$ and $α_4 < 0$, the direction of transit orbits is reversed. Figure 7(c) and (d) presents the non-transit orbits of halo orbits and quasihalo orbits with $α_3α_4 > 0$. When $α_3 < 0$ and $α_4 < 0$, non-transit orbits are restricted to the left side of the halo/quasihalo orbit. When $α_3 < 0$ and $α_4 < 0$, they are restricted to the right side.

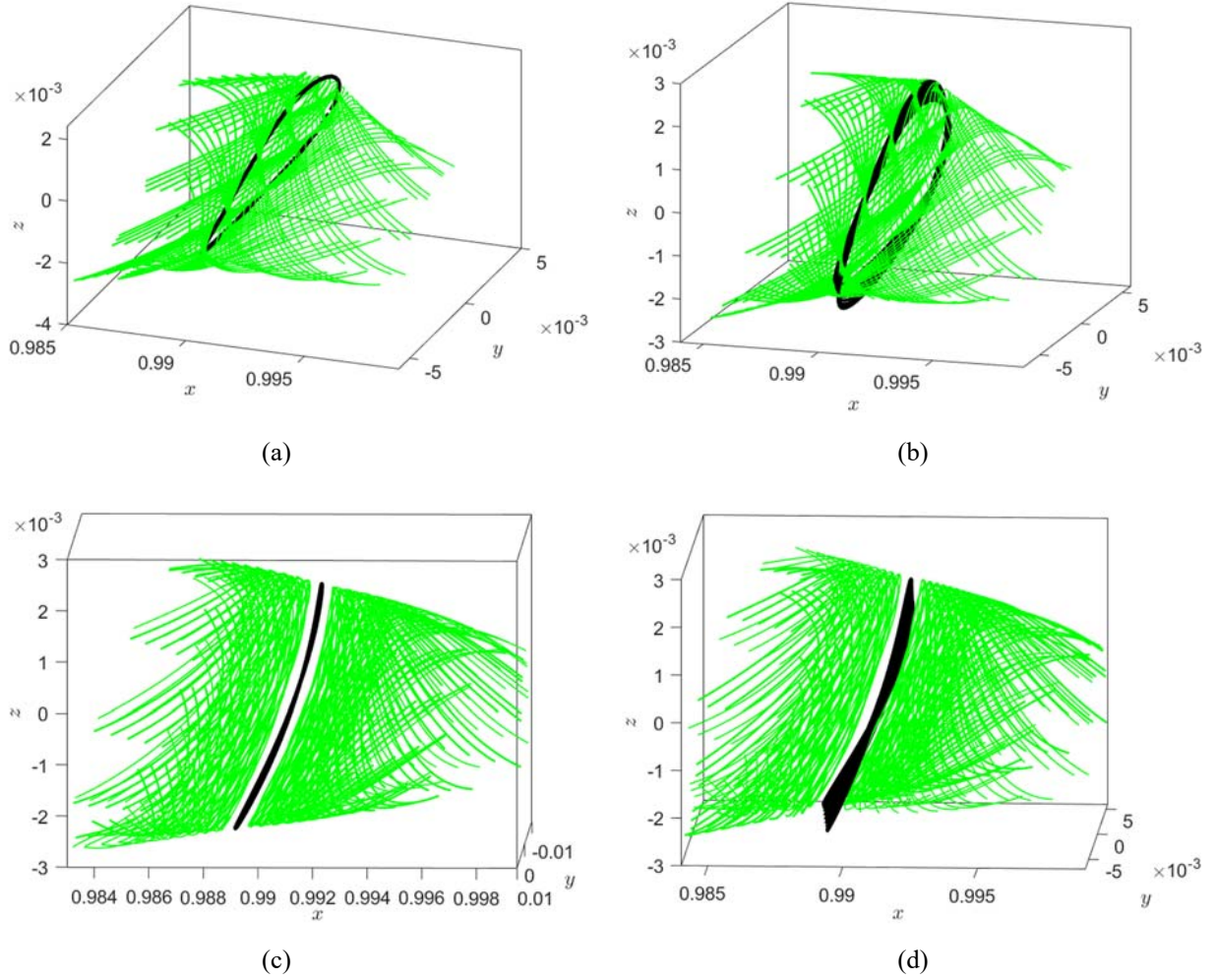

Fig. 7 (a) Transit orbits with $α_1 = 0.16$, $α_2 = 0$. (b) Transit orbits with $α_1 = 0.16$, $α_2 = 0.02$. (c) Non-transit orbits with $α_1 = 0.16$, $α_2 = 0.02$. (d) Non-transit orbits with $α_1 = 0.16$, $α_2 = 0.02$.

Finally, we analyze the accuracy of the series solution obtained using the coupling-induced bifurcation mechanism by comparing it with high-precision numerical solutions. When the series solution only describes the center manifold, it is consistent with the findings in [17]. To establish a good benchmark for comparison, we analyze the accuracy of the solution using the invariant manifolds corresponding to Lissajous orbits and Quasihalo orbits as examples. Like [16], we choose $α_3 = 0.001$, $α_4 = 0$ (for unstable manifolds), $\varphi_1 = \varphi_2 = 0$. Then, after specifying any pair of ($α_1$, $α_2$) values, initial values can be obtained using the series solution. Subsequently, the numerical integration of the system equations is performed until the error between the analytical solution and the numerical solution reaches $10^{-6}$.

Figure 8 describes the convergence of the 15th-order analytical solution when the coupling coefficient is zero and non-zero. The time span of numerical integration in Fig. 8 represents the degree of agreement between the numerical solution and the series solution up to order 9 and 15. It can be observed that when the $\eta = 0$, corresponding to the unstable manifold associated with Lissajous orbits, the results align with those reported in [16]. However, the actual motion amplitude in the direction of $\alpha_2$ is a superposition of the effects of motion coupling in the direction of $\alpha_1$. This results in a significantly smaller convergence region for the unstable manifold corresponding to quasihalo orbits compared to Lissajous orbits, as shown in Figure 8(b).

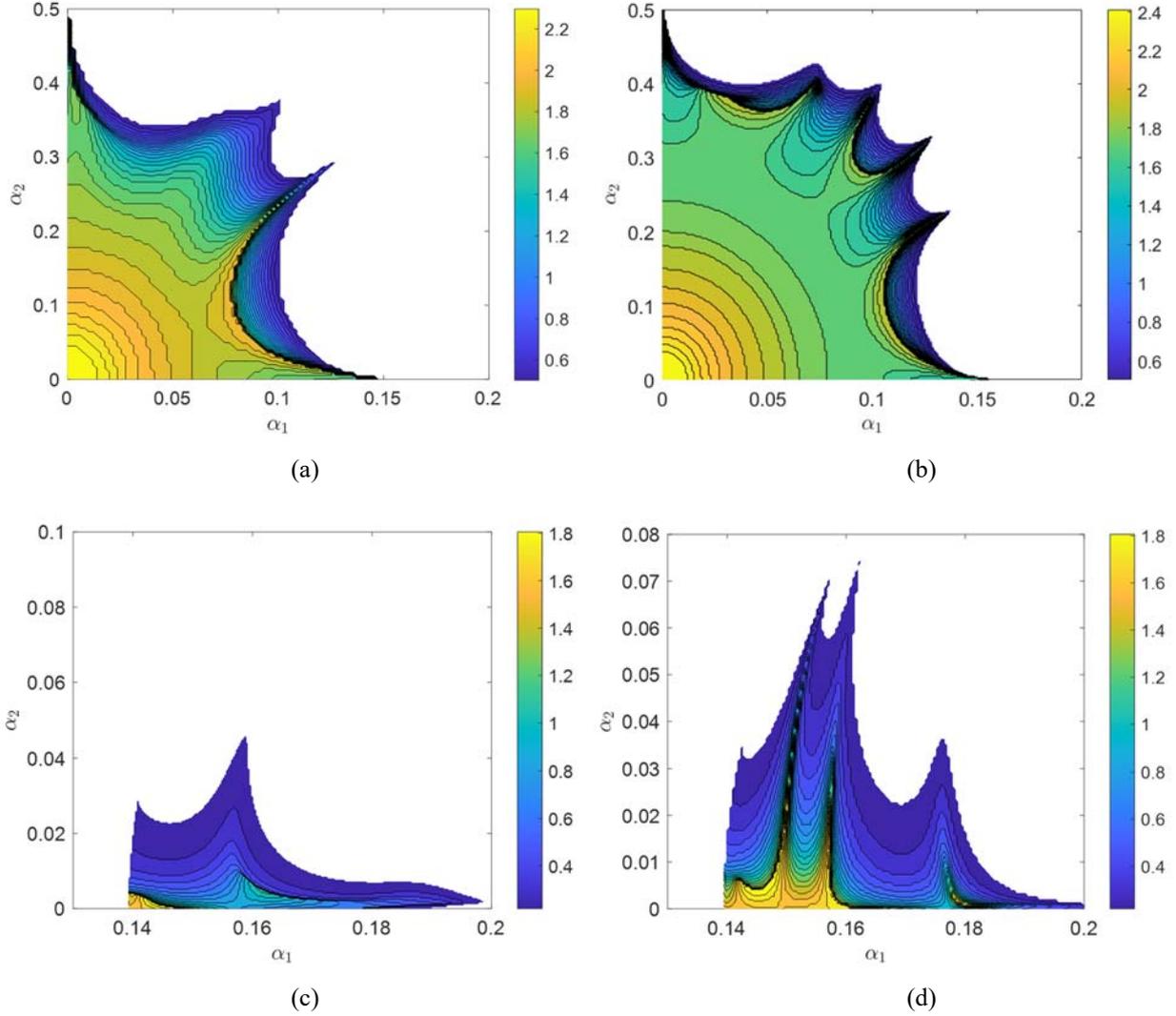

Fig. 8 Time span when the error between the proposed analytical solution and the numerical solution reaches $1\times10^{-6}$. (a) Lissajous orbit's unstable manifolds $\eta = 0$, $n = 9$. (a) Lissajous orbit's unstable manifolds $\eta = 0$, $n = 15$. (c) Quasihalo orbit's unstable manifolds, $\eta \neq 0$, $n = 9$. (c) Quasihalo orbit's unstable manifolds, $\eta \neq 0$, $n = 15$.

## 5. Conclusion

This paper presents a unified analytical solution that comprehensively describes the phase space near collinear libration points in the CRTBP using the coupling-induced bifurcation mechanism. The analytical framework (16) encompasses Lissajous orbits, quasihalo orbits, their invariant manifolds, and transit and non-transit orbits. Specifically, for the first time, an approximate analytical solution for the invariant manifold corresponding to quasihalo orbits is provided, which has not been achievable by other methods thus far. The introduction of a coupling coefficient $\eta$ and a bifurcation equation $\Delta = 0$ is pivotal in systematically deriving a uniform series solution for these orbits, achieved through the Lindstedt-Poincaré method. Additionally, the bifurcation equation $\Delta = 0$ transforms the dynamical bifurcation problem of orbits near libration points into a static bifurcation problem of polynomial equation, greatly simplifying the bifurcation analysis of the CRTBP. By analyzing the third-order bifurcation equation (25), the critical bifurcation conditions for halo orbits, quasihalo orbits, and their corresponding invariant manifolds is elucidated. Notably, the analysis also unveils two new families of periodic orbits akin to classical halo orbits. Furthermore, not only invariant tori but also general transit and non-transit orbits in hyperbolic subspaces undergo bifurcations.

The derived bifurcation equation (25) indicates that the coupling coefficient $\eta$ is only related to the parameters $\alpha_1^2$, $\alpha_2^2$, and $\alpha_3\alpha_4$, specifically, $\eta = \eta(\alpha_1^2, \alpha_2^2, \alpha_3\alpha_4)$. This implies that each parameter corresponds to one degree of freedom in the system. Therefore, for a general $n$-dimensional Hamiltonian system, its bifurcation equation will involve $n$ parameters.

In sum up, the proposed unified analytical framework provides a holistic view of the phase space structures near collinear libration points in the CRTBP. It also addresses analytical challenges related to quasihalo orbits and their invariant manifolds, enhancing our understanding of celestial mechanics and potentially influencing the design and analysis of space missions.